\documentclass[aps,axodraw,twocolumn,prl,nofootinbib,superscriptaddress]{revtex4}
\usepackage{fancyhdr}
\usepackage{fancybox}
\usepackage{epstopdf}

\usepackage{graphicx}
\usepackage{color}
\usepackage{soul}
\usepackage{latexsym}

\begin{document}

\preprint{UCD-2008-01}
\preprint{CERN-PH-TH/2008-145}
\title{\boldmath 
Signatures of Extra Dimensions from Upsilon Decays \\ with a Light Gaugephobic Higgs Boson}

\author{Jamison Galloway}

\affiliation{Department of Physics, University of California, Davis, CA 95616}

\author{Bob McElrath}

\affiliation{CERN, Geneva 23, Switzerland}

\author{John McRaven}

\affiliation{Department of Physics, University of California, Davis, CA 95616}

\date{\today}

\begin{abstract}
We explore non-standard Higgs phenomenology in the Gaugephobic Higgs model in
which the Higgs can be lighter than the usually quoted current experimental
bound.  The Higgs propagates in the bulk of a 5D space-time and Electroweak
Symmetry Breaking occurs by a combination of boundary conditions in the extra
dimension and an elementary Higgs.  The Higgs can thus have a significantly
suppressed coupling to the other Standard Model  fields.  A large enough 
suppression can be found
to escape all limits and allow for a Higgs of any mass, which would be
associated with the discovery of $W^\prime$ and $Z^\prime$ Kaluza-Klein
resonances at the LHC.  The Higgs can be
precisely discovered at B-factories while the LHC would be insensitive
to it due to
high backgrounds.  In this letter we study the Higgs discovery mode in
$\Upsilon(3S)$, $\Upsilon(2S)$, and $\Upsilon(1S)$ decays, and the model
parameter space that will be probed by BaBar, Belle, and CLEO data.  In
the absence of an early discovery of a heavy Higgs at the LHC, A Super-B
factory would be an excellent option to further probe this region.

\end{abstract}


\maketitle

\section{Introduction} 
If electroweak symmetry breaking in the Standard Model (SM) arises solely from
the presence of a fundamental scalar, the scale of the electroweak interactions
requires a severe fine-tuning.  The economy of the Higgs mechanism thus comes
at the cost of making the SM unnatural.  Technicolor models \cite{TC} aim to
ameliorate this instability by considering the Higgs as a composite
state; however, these simplest models are ruled out  by their large
oblique corrections \cite{TCoblique}.  A new approach to a composite
Higgs is provided by the AdS/CFT correspondence
\cite{AdS/CFT}, in particular as represented by Randall-Sundrum (RS1)-type
setups \cite{RS1}.  Typically the Higgs has been confined to a
particular brane in the 5D picture, thus corresponding to a 4D state of
infinite scaling dimension \cite{dimension}. This, however, is more than
is necessary to avoid issues of extreme fine-tuning.  Even if the Higgs is
localized somewhere near the IR brane of RS1, the corresponding 4D state is
interpreted as a composite and can be light with tuning at only the percent
level. 
This particular
relaxation of the usual assumptions is the salient feature of the Gaugephobic
Higgs model \cite{Gaugephobic} we consider below (see also \cite{bulkhiggs} for
other treatments of a 5D Higgs).  The crucial aspect of this model that we
exploit is that the Higgs can be made light (e.g. $m_H  < 10$ GeV) while
simultaneously suppressing its couplings to fermions and weak gauge bosons,
such that current experimental constraints are evaded.

\section{The Gaugephobic Higgs Model}
The Gaugephobic model is described in \cite{Gaugephobic}; here we 
review only the features important for Higgs production at B-factories.  As in
RS1, we have a slice of AdS$_5$ with conformally flat metric (taking
$z$ to denote the coordinate of the extra spatial dimension):
\begin{equation}
ds^2 = \left(\frac{R}{z}\right)^2 (\eta_{\mu \nu} dx^\mu dx^\nu-dz^2).
\end{equation}
$R$ corresponds to the position of the UV brane and sets the curvature scale of
the extra dimension.  The second boundary is at $z=R'$ with $R' \gg R$
generating the weak-Planck hierarchy due to the warp factor.  $R$ is a free
parameter, while $R'$ is set by the masses of the weak gauge bosons.  The bulk
gauge group $SU(2)_L \times SU(2)_R \times U(1)_X$ is broken to
$U(1)_{EM}$ by boundary conditions and a bi-fundamental Higgs 
with zero $X$ charge.  With the Higgs taken to be a bulk field, we choose the three
parameters $\beta, m_H, V$ to describe it.  In our analysis we parameterize the
effect of the Higgs bulk mass $\mu$ by $\beta \equiv \sqrt{4+\mu^2}$.
Conventional RS1 is described by the limit $\beta \to \infty$.

The profile of the vacuum expectation value (VEV) is
controlled by UV brane boundary conditions to be 
\begin{equation}
\label{VEVprofile}
    v(z) = \sqrt{\frac{2(1+\beta)\log R'/R}{1-(R/R')^{2(1+\beta)}}}
           \frac{g V}{g_5} \frac{R^\prime}{R} \left(\frac{z}{R^\prime} \right)^{2+\beta},
\end{equation}
where $g$ is the SM $SU(2)$ gauge coupling, and $g_5$ is the 5-dimensional 
$SU(2)_{L/R}$ gauge
coupling.  The normalization $V$ of the VEV is chosen such that the SM is 
recovered as one takes $V\to 246$ GeV: in this limit the gauge 
boson profiles are flat, with all mass coming from direct overlap with the Higgs.  
Conversely, in the limit $V\to  \infty$ the profiles 
of the gauge bosons are pushed towards the UV (away from the IR-localized VEV) 
so that their mass comes entirely from momentum in the fifth dimension.  
This corresponds to the Higgsless limit \cite{higgsless}: in this case the 
Kaluza-Klein (KK) scale is lowered, so that the appearance of the weakly-coupled KK states 
fulfill the Higgs boson's additional role of restoring unitarity in $WW$-scattering. 


The other ingredient that establishes the profile (\ref{VEVprofile}) is
the Higgs quartic coupling $\lambda$, which is confined to the IR brane
to ensure that electroweak symmetry breaking takes place there.  We
trade this parameter for the mass $m_H$ of the physical Higgs mode via
the effective potential's minimization condition, in the same way as in
the SM.  The couplings between the Higgs and other states is provided by
the overlap of the corresponding 5D profiles, so field localization
governs interaction strength.  

\begin{table}[t]
    \begin{center} 
        \begin{tabular}{c|c}
            Parameter  & Range \\ \hline 
            $m_h$ [GeV] & [0, \ 10] \\ \hline
            $\beta$ & [2, \ 10] \\ \hline 
            $V$ [GeV] & [250, \ 1500] \\ \hline
            $c_L$($b$) & [0, \ 0.5] \\ \hline 
            $c_R$($b$) & [-0.79, \ -0.7]
        \end{tabular}
    \caption{Range of the scanned parameter space with the AdS scale set
    by $R^{-1}=10^8$ GeV.  The range of $\beta$ is chosen to localize
    the Higgs VEV towards the IR brane, while the range of $V$ is chosen
    to interpolate between the SM and ``almost Higgsless'' limits.  The
    bulk mass for the left- and right-handed bottom quark are
    constrained by the required precision of their coupling to the $Z$.}
    \label{tab:params} 
    \end{center} 
\end{table}

The light fermions in the model are arranged in doublets of the bulk
gauge group.  The 5D fermions must be 
vector-like due to the nature of the 5D realization of the Dirac algebra, 
so that bulk mass terms are allowed for them and will dictate their localization.
They each have dimensionless bulk
masses $c_L$ and $c_R$ for the left- and right-handed pieces as well as
a UV kinetic term to split the masses within a given multiplet.
The inclusion of the third
quark generation requires more care, however, since the heavy top quark
requires a large overlap with the Higgs VEV.  With the top and bottom 
arranged together in doublets, this would lead to an unacceptable deviation in
the $Zb_L {\bar b}_L$ coupling.  We choose to solve this problem as in
\cite{custodian} where non-universal corrections to the $Z$-couplings
are avoided by representing the left-handed bottom quark in a bi-doublet
of the bulk $SU(2)_L \times SU(2)_R$.  The total field content of the
third generation thus contains the new fields $T$ and $X$,
where the quantum numbers of the $T$ allow it to mix with $t$.
The new exotic quark $X$ has electric charge 5/3 so won't mix with
the other fields.  The lowest lying $X$ state enters at
$m_X \sim 1 \ {\rm TeV}$.



\section{Parameter Space and Constraints}

\begin{figure}[t]
    \centering
    \includegraphics[width=8.4cm]{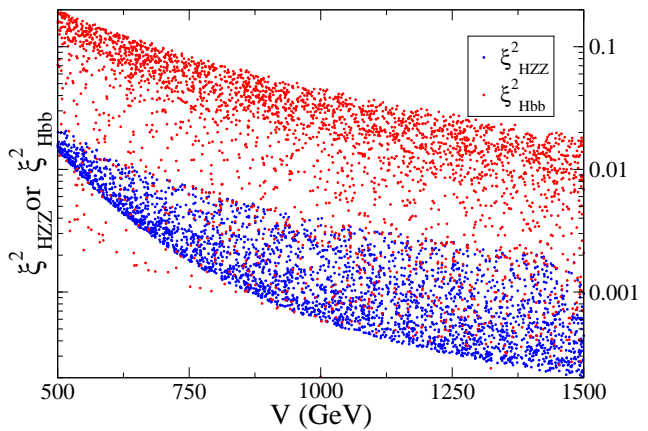}
    \hspace{0pt} 
    \caption{$\xi^2$ vs. $V$.  As $V\rightarrow 246$ GeV from above the
    SM is approached, i.e. $g_{HZZ}\rightarrow g_{HZZ}^{\rm SM}$ while
    as $V$ is increased the gauge bosons decouple from the Higgs.} 
    \label{fig:HZZHbbvsV}
\end{figure} 

The Gaugephobic model is described by the five parameters shown in
Table~\ref{tab:params}, with the ranges we considered.  In
Fig.~\ref{fig:HZZHbbvsV} 
we scan over
the parameter space imposing the constraints in this section.  We find
that {\it all} of the Higgs couplings are suppressed in this
model.

LEP searched for the Higgs in the Higgsstrahlung mode in which it is
radiated off a $Z$ boson through the $HZZ$ coupling.  By decoupling the
Higgs from the $Z$, LEP would have a sufficiently small rate that it
could not discover the Higgs~\cite{LEP}. We apply the decay mode independent bound on the
Higgsstrahlung cross section.  This limit varies by a factor of two as a function of mass; we apply $\xi_{HZZ}^2 < 2.1 \times 10^{-2}$ which is the upper bound for the limit in the range $2 m_\tau < m_H < m_{\Upsilon(3S)}$, where
we define the suppression relative to the SM of $Z$ bosons and bottom
quarks as 
\begin{equation}
    \xi^2_{HZZ} \equiv \left(g_{HZZ}/g_{HZZ}^{SM}\right)^2;
    \qquad
    \xi_{bbH}^2 \equiv \left(y_b/y_b^{SM}\right)^2,
    \label{eq:xiHZZ}
\end{equation}
with $g_{HZZ}$  
denoting the $H \to ZZ$ coupling and $y_b$ the bottom Yukawa.   These
suppression factors are shown in
Fig.~\ref{fig:HZZHbbvsV}
and are uncorrelated with
the Higgs mass.  The LEP constraint depends only on the
$HZZ$ coupling and is independent of other modifications which would
change the Higgs decays.


With the Higgs decoupled from the $Z$, the next most relevant
constraints come from radiating the Higgs off $b$ quarks.  For $2 m_\mu
< m_H < 2 m_\tau$, the SM Higgs was first ruled out by
ARGUS~\cite{Alam:1989mta} in the channels $B \to  K H$ and $B \to K^* H$ 
with the assumption that $m_t = 50$ GeV.  However today we know from CDF
and D0~\cite{Yao:2006px} that $m_t = 172$ GeV, which strongly enhances this branching
ratio.  For a SM Higgs in this mass range, these channels
would be dominant~\cite{Grinstein:1988yu} because of an $m_t^4$
enhancement in the rate:
\begin{eqnarray}
    \label{eq:GammabHs}
    \frac{\Gamma(b \to H s)}{\Gamma(b \to c e \overline{\nu}_e)} = 
    \qquad \qquad \qquad \qquad \qquad \qquad \qquad 
    \\
    \frac{27 \sqrt{2}}{64 \pi^2} G_F m_b^2
    \frac{\left(1-\frac{m_H^2}{m_b^2}\right)^2}{f(m_c/m_b)} \left|\frac{V_{st}^\dagger
    V_{tb}}{V_{cb}}\right|^2\left(\frac{m_t}{m_b}\right)^4,
    \nonumber
\end{eqnarray}
where $f(m_c/m_b)\sim 0.5$ is the dimensionless phase space factor for
$b \to ce\overline{\nu}_e$.  We use this standard result to approximate
the rate even in this model.  New contributions coming from KK quarks
will contain suppression not only from the top Yukawa couplings, but also
from both gauge couplings appearing in the diagram: the overall
suppression from these three couplings makes their contribution
substantially smaller than Eq.~\ref{eq:GammabHs}.  The exotic $X$ quark
does not contribute to this process.
Thus to avoid regions that are tightly constrained to have an extremely
weak Higgs coupling, we prefer $m_H > 2 m_\tau$.  However, as can be seen
in Fig.~\ref{fig:HZZHbbvsV}, the
couplings of the Higgs become arbitrarily small as $V \to \infty$, so that a 
large enough VEV could  provide an adequate suppression in the top Yukawa 
coupling to explain the observed rate. With the measured value~\cite{Yao:2006px} 
of $B \to s \mu^+ \mu^-$ and assuming $BR(H\to \mu^+ \mu^-) = 5\%$, 
the Gaugephobic Higgs with $m_H<2 m_\tau$ is allowed when 
$V
> 3.1$ TeV.
At this point we have a suppression of the top Yukawa coupling
$\xi^2_{ttH} \sim 10^{-5}$ while $\xi^2_{bbH} \sim 10^{-4}$.

%

For $m_H > 2 m_\tau$ the most profitable mode to search is in
$\Upsilon(nS) \to \gamma H$~\cite{Wilczek} where $n = 1,2,3$, which we discuss in
detail in the next section.  Once the $HZZ$ constraints
are taken into account, the Gaugephobic Higgs also has suppressed
couplings to $b$ quarks and therefore $\Upsilon$'s.  This mode was not
as vigorously pursued as Higgsstrahlung and $B$ meson decays because
there is sufficient theoretical uncertainty in the predictions for this
mode.  Even including these uncertainties, this mode only barely reached
the expected SM level.  Therefore LEP data was used to rule out the SM
Higgs in the $m_B-m_K < m_H < M_\Upsilon$ region instead.  Searches were
performed by the CLEO collaboration using $\Upsilon(1S)$ decays to
mono-energetic photons~\cite{Besson:1985xw}.  They limit \[ BR(\Upsilon(1S)
\to \gamma H) < 0.4\%; \qquad 8.4 {\rm GeV} < M_H < 9.4 {\rm GeV}. \]

The CUSB Collaboration measured the entire photon spectrum from Upsilon
decays~\cite{Franzini:1987pv}.  They rule out earlier claims from 
Mark III~\cite{Baltrusaitis:1985pu} and
Crystal Ball~\cite{xiclaim} of evidence for Higgs resonances at 2.2 GeV
and 8.3 GeV respectively.  
This limit just barely reaches the SM expectation
$BR(\Upsilon \to \gamma H) \sim 2 \times 10^{-4}$ for $M_H \to 0$ and
worsens to limit $BR(\Upsilon \to \gamma H) < 1.5 \times 10^{-3}$ as
$M_H$ increases.  

Finally the ARGUS collaboration searched for a monochromatic photon
line~\cite{Albrecht:1985qz} in the ranges
\begin{eqnarray}
    \nonumber
    BR(\Upsilon(1S) \to \gamma H) < 0.1\%;&& \quad 2.1{\rm GeV} < m_H < 8.9{\rm GeV} \\
    \nonumber
    BR(\Upsilon(2S) \to \gamma H) < 0.5\%;&& \quad 3.2{\rm GeV} < m_H < 9.5{\rm GeV}
    \label{eq:ARGUSlimits}
\end{eqnarray}
where the limits quoted are at the lowest $m_H$ and worsen slightly for
higher $m_H$.

Additionally, there is an important indirect constraint from the coupling of
the $Z$ to $b$ quarks, $g_{Zbb}$: for left-handed $b$'s this is
constrained to be within $\sim$0.25\% of its SM value \cite{custodian}
while for the right-handed fields the constraint is relaxed to
$\sim$30\%~\cite{Choudhury:2001hs}.  This accuracy is possible only with
the third generation incorporated in the representations described
above, and even then provides a stringent condition on the bulk masses
of those fields.

We point out that a complete analysis of electroweak precision
parameters is lacking for this model.  However it has been shown that in
the Higgsless limit, the large contributions to the $S$-parameter
typical of Technicolor models can in fact be cancelled in a holographic
model by an appropriate ``de-localization'' (i.e. tuning of the bulk
masses) of the bulk fermions \cite{delocalization}.  The effect of
de-localization on our results is small: we have confirmed  numerically
that adding restrictions to the localization of the light fermions does
not qualitatively change our results. 

\section{A Light Higgs in $\Upsilon$ Decays} 


At low masses, the Gaugephobic Higgs is produced by radiation from the
heaviest fermion available.  Data with heavy fermions comes dominantly
from producing $\Upsilon$ and $J/\Psi$ resonances.  BaBar has collected
30.2 fb$^{-1}$ on the $\Upsilon(3S)$ and 14.45 fb$^{-1}$ on the
$\Upsilon(2S)$, complementing the 3 fb$^{-1}$ collected by Belle, and
older results from CLEO.

\begin{figure}[t]
    \centering
    \includegraphics[width=8cm]{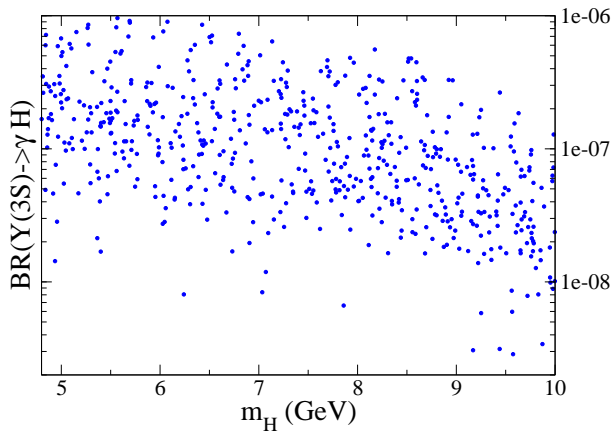}
    \hspace{0pt} 
    \caption{Branching ratio of the $\Upsilon(3S)$ to a photon and
    Higgs, as a function of Higgs mass.}
    \label{fig:BR}
\end{figure}

The Higgs is radiated from vector resonances $V \to \gamma
H$~\cite{Wilczek}.  The photon is monochromatic with an energy
\begin{equation}
    E_\gamma = \frac{M_V^2-M_H^2}{2 M_V}
    \label{eq:egamma}
\end{equation}
because the Higgs is extremely narrow ($\Gamma_H < 1$ MeV) for these
masses.  
The relative rate assuming a Coulomb-like potential for the $b {\bar b}$
state is~\cite{Wilczek}
\begin{eqnarray}
    \frac{\Gamma(\Upsilon \to H \gamma)}{\Gamma (\Upsilon \to \mu \mu)} &=&
    \frac{G_F \ m_b^2}{\sqrt{2} \pi \alpha} \left(1-\frac{m_h^2}{m_\Upsilon^2}\right)\, 
    \xi_{Hbb}^2 \epsilon \, ; \\
    BR(\Upsilon \to H \gamma) &\simeq& 
    1 \times 10^{-4} \left( 1-\frac{m_H^2}{m_\Upsilon^2}\right)\, \xi_{Hbb}^2\epsilon \, ,
\end{eqnarray}
where $\xi_{Hbb}$ is the suppression relative to 
the SM.  The factor $\epsilon$ includes any next-to-leading order
corrections, most notably the leading one-loop QCD
correction~\cite{Vysotsky:1980cz,Barbieri:1975ki,Nason:1986tr}
and relativistic correction~\cite{aznauryan}.
All of these corrections reduce the branching ratio to
Higgs over the entire mass range, but there is considerable uncertainty as
to how to combine the various contributions.  See \cite{HHG} for further
discussion.    Since these two corrections are coming respectively from
hard and soft gluon effects, we simply combine the two to find the
approximate branching fraction for $\Upsilon (3S) \to H \gamma$  shown
in Fig.~\ref{fig:BR}.  The relative uniformity of this plot
reflects the fact that the suppression of the bottom Yukawa coupling has
little direct dependence on the mass of the physical Higgs.  Numerical differences
between this rate for the $3S$ state and the same rate for the lighter
$n=1,2$ resonances can be determined from the difference in the partial
width $\Gamma(\Upsilon \to \mu \mu)$ of each.

Unfortunately the $\Upsilon(4S)$ data is almost useless in the Wilczek
mode because its width is so much larger.  For the $\Upsilon(4S)$ data
to be competitive with $\Upsilon(3S)$ data, one needs approximately
$\Gamma_{\Upsilon(4S)}/\Gamma_{\Upsilon(3S)} \simeq 1000$ times more
data because the $\Upsilon(4S)$ is above threshold for decay into a pair
of $B$ mesons and consequently has a very large width.  However, one can
profitably search for a Higgs in $B$ meson decays using $\Upsilon(4S)$
decays, albeit with reduced kinematic reach $m_H < 4.8$ GeV.

\section{Conclusions}  
A light Higgs boson is experimentally excluded only when its couplings
to other SM fields are sufficiently large.  There still exists a class of
viable models in which these couplings are suppressed in an ``almost
Higgsless'' scenario, allowing for the potential discovery of a light
Higgs at B-Factories.  This discovery would be associated with the
discovery at the LHC of heavy $Z^\prime$ and $W^\prime$ Kaluza-Klein
resonances and no Higgs.  We show the range of viable parameters
within the Gaugephobic Higgs model.  For a Higgs lighter than 10 GeV,
the relevant signal would be an excess of monochromatic photons in
$\Upsilon(nS)$ data, associated with a pair of heavy fermions such as
charm or tau.  A Higgs lighter than the $B$ meson is much more tightly
constrained to be nearly Higgsless, and can be discovered in $B \to K H$
using $\Upsilon(4S)$ data.

\section{Acknowledgements} 
We thank Christophe Grojean, Jack Gunion, Damien
Martin, and John Terning for discussions.  
The work 
of J.G. and J.M. is supported by the US Department of Energy under 
contract DE-FG03-91ER40674.


\end{document}